\documentclass[twocolumn,preprintnumbers,amsmath,amssymb]{revtex4}
\usepackage{graphicx}
\usepackage{dcolumn}
\usepackage{bm}
\usepackage{comment}
\usepackage{rotating}
\usepackage{longtable}
\usepackage{float}
\usepackage{eucal}
\usepackage{csquotes} 
\usepackage{multirow}
\usepackage{booktabs}



\begin{document}

\title{Coupling between two extreme excitation mode in weakly deformed $^{142}$Eu nucleus}
\author{Sajad Ali$^{1}$}
\author{S. Rajbanshi$^{1,2}$}
\author{Prithwijita Ray$^{1}$}
\author{Somnath Nag$^{3}$}
\author{Abhijit Bisoi$^{4}$}
\author{S. Saha$^{5}$}
\author{J. Sethi$^{6}$}
\author{T. Trivedi$^{7}$}
\author{T. Bhattacharjee$^{8}$}
\author{S. Bhattacharyya$^{8}$}
\author{S. Chattopadhyay$^{1}$}
\author{G. Gangopadhyay$^{9}$}
\author{G. Mukherjee$^{8}$}
\author{R. Palit$^{10}$}
\author{R. Raut$^{11}$}
\author{M. Saha Sarkar$^{1}$}
\author{A. K. Singh$^{12}$}
\author{A. Goswami$^{1}$}
\email{asimananda.goswami@saha.ac.in}

\affiliation{$^1$Saha Institute of Nuclear Physics, HBNI, 1/AF, Bidhannagar, Kolkata 700064, India}
\affiliation{$^2$Dum Dum Motijheel College, Kolkata 700074, India}
\affiliation{$^3$IIT-BHU, Banaras Hindu University Campus, Varanasi 221005, India}
\affiliation{$^4$Indian Institute of Engineering Science and Technology, Howrah 711103, India}
\affiliation{$^5$GSI Helmholtzzentrum für Schwerionenforschung, Darmstadt, Hesse, Germany}
\affiliation{$^6$Department of Chemistry and Biochemistry, University of Maryland, United States}
\affiliation{$^7$Guru Ghasidas Vishayavidyalaya, Bilaspur 495009, India}
\affiliation{$^8$Variable Energy Cyclotron Center, Kolkata 700064, India}
\affiliation{$^9$Department of Physics, University of Calcutta, Kolkata 700009, India}
\affiliation{$^{10}$Tata Institute of Fundamental Research, Mumbai 400005, India}
\affiliation{$^{11}$UGC-DAE-Consortium for Scientific Research, Kolkata 700098, India} 
\affiliation{$^{12}$Indian Institute of Technology, Kharagpur 721302, India}

\date{\today}

\begin{abstract}

Two opposite parity dipole bandlike structures DB I and DB II of $^{142}$Eu are investigated by the Indian National Gamma Array (INGA), using the fusion evaporation reaction $^{31}$P + $^{116}$Cd @ 148 MeV. The decreasing trend as well as magnitude of the measured $B(M1)$ and $B(E2)$ transition rates of the band DB II has been reproduced well within the shears mechanism with the principal axis cranking model calculations. This calculation reflects the fact that the maximum contribution of the angular momentum of the states in DB II has been generated from the magnetic rotation (MR) phenomenon. The enhanced $B(E1)$ rates of the connecting $E1$ transitions from the states of DB II to DB I are demanding the octupole correlation due to the involvement of the octupole driving pair of orbitals $\pi{h_{11/2}}$ and $\pi{d_{5/2}}$ as evident from the quasiparticle alignment ($i_{x}$), the experimental routhians (e$^{'}$) and the calculated neutron and proton quasiparticle energies against the rotational frequency ($\omega$). 

\end{abstract}

\maketitle

Generation of angular momentum in weakly deformed nuclei is long standing debate. These nuclei are expected to be spherical at lower excitation energies and spin which can be interpreted quantitatively on the basis of spherical shell model calculations. At higher excitation energies, multi particle configuration along with its small deformation plays an important role in determining the level structures for these nuclei. Transitional nuclei in mass A $\sim$ 140 region are crucial laboratory to observe the interesting nuclear structure phenomena and to test variety of nuclear models. Due to the proximity of the spherical shell closures and competing shape (prolate and oblate) driving effects of the high-j orbital near the proton and neutron Fermi levels several novel phenomena, like shape co-existence, shears mechanism, octupole correlation, chiral symmetry breaking etc are expected in the excited spectrum of these nuclei \cite{fraun2, mac, clark}. 

Among these magnetic rotation is more common phenomena for generating angular momnetum in weakly deformed nuclei. During the last three decades, regular sequence of rotational-like bands consisting of stretched dipole transitions (M1) have been observed for several weakly deformed nuclei near shell closure in different mass regions and proposed as shears mechanism \cite{mac, clark}. These bands came to be known as magnetic rotational (MR) band in which the high-$j$ valence particles and holes coupled to generates angular momentum. 

Breaking of rotational symmetry results the rather common phenomena quadrupole deformation. Octupole moment is the next relevant after quadrupole and it characterized by the breaking of the reflection symmetry under space inversion. With respect to the quadrupole deformation the number of permanent octupole deformation is very few. Though, interaction between the orbitals of near the Fermi surface which differ by three units of angular momenta, leading to a sequence of interleaved states of opposite parity. Typically, this situation occurs when Fermi level is found between intruder sub-shell ($l, j$) and normal parity sub-shell ($l-3, j-3$) \cite{rev}. The dynamical correlation between this sub-shell results the connection of enhanced $E1$ transitions between the sequences of interleaved states of negative and positive parity called octupole correlation.

The weakly deformed nuclei in mass $\sim$ 140 region has vacancy of proton in $d_{5/2}$ orbital and vacancy of neutron in the intruder orbital $h_{11/2}$. The neutron holes are restricted to this orbital, but the proton can be easily excited to the intruder orbital $h_{11/2}$ across the Z = 64 sub-shell closure. The interaction between $d_{5/2}$ and $h_{11/2}$ orbitals of the proton sector near the Fermi surface may leads the band of interleaved states of negative and positive parity connected by $E1$ transitions. Also, the presence of particles and holes in high $j$ orbital magnetic rotational band can also be observed as reported in other nuclei in this mass region \cite{139sm, sajad, rajban, pod, rajban2, past}. This opens up the possibility of observing of two extreme mode of excitation mechanism in the vicinity of $Z = 64$ sub-shell closure. The $^{142}$Eu nucleus has one proton hole in $d_{5/2}$ and three neutron holes in $h_{11/2}$ orbital is on of the possible candidate for observing above mentioned characteristics. So, in this mass region $^{142}$Eu is studied for the quest of observing the coupling between two extreme excitation mode which may leads connection between two modes.


The structure of interest in $^{142}$Eu nucleus has been populated through the fusion evaporation reaction of the $^{31}$P as projectile with the lead (14.5 mg/cm$^{2}$) backed $^{116}$Cd (99$\%$ enriched) target having thickness 2.4 mg/cm$^{2}$. The 148 MeV $^{31}$P beam was obtained from the 14UD Pelletron Linac facility at TIFR, Mumbai. The purpose of using the Pb baking was to ensure that all residual nuclei, moves with initial velocity $v_{0}$ $\sim$ 2\% of c, populated in this reaction have been stopped inside the target-backing combination. The de-exciting $\gamma$ rays were detected by the Indian National Gamma Array (INGA) which was consisted of nineteen Compton-suppressed clover detectors at the time of experiment. The detectors were arranged in six different rings, with four detectors at 90$^{\circ}$ and each three at 40$^{\circ}$, 65$^{\circ}$, 115$^{\circ}$, 140$^{\circ}$, and 157$^{\circ}$ with respect to the beam direction.

In offline mode, the time-stamped data has been sorted into usual coincident events based on the mapping of the digital data acquisition (DDAQ) channels to different crystals of the detectors of the INGA set up \cite{palit, htan} using the multi-parameter time-stamped based coincidence search program (MARCOS) developed at the TIFR, Mumbai. The MARCOS program was used to find out the energy and efficiency calibrations of the clover detectors in the array in which the target has been replaced by the radioactive sources $^{152}$Eu and $^{133}$Ba. The collected two and higher-fold coincidence events (4.0 $\times$ 10$^{9}$) were sorted into several symmetric and asymmetric $E_\gamma$ - $E_\gamma$  matrices and $E_\gamma$ - $E_\gamma$ - $E_\gamma$ cube which were analyzed subsequently with the help of the RADWARE and the INGASORT software packages \cite{radford1, radford2, ingasort}. Details of the experiment and data-analysis procedures have been described in Refs. \cite{rajban, rajban2, sajad}.

\begin{figure}[t]
\centering
\setlength{\unitlength}{0.05\textwidth}
\begin{picture}(10,8.8)
\put(-1.8,9.2){\includegraphics[width=0.50\textwidth, angle=-90]{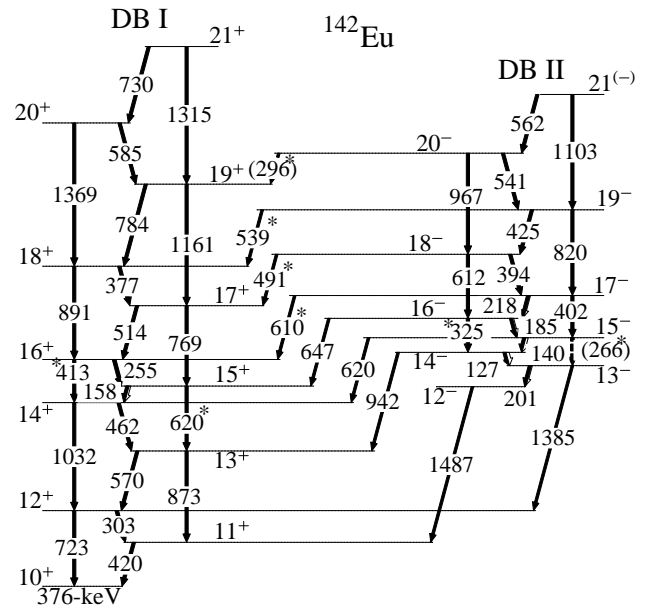}}
\end{picture}
\caption{\footnotesize The partial level scheme of $^{142}$Eu obtained in the present work. The $\gamma$ -ray energies are rounded off to the nearest keV. The newly observed $\gamma$ -ray transitions are marked by an asterisk.}
	\label{levelsc}
\end{figure}

\begin{table*}
   \centering
\caption{{\label{table1} Measured level lifetimes, branching ratios, reduced transition probabilities ($B(M1)$, $B(E1)$ and $B(E2)$ values) and the $B(E1)/B(E2)$ ratios for the excited states in the negative parity bands in \text{$^{142}$Eu}.}}
\begin{ruledtabular}
\begin{tabular}{cccccccccccc}
\midrule
\midrule

\text{$J^\pi$} & \text{$E_\gamma$}$^{a}$  & \text{$R_{DCO}$$^{a}$}  & \text{$R_{\theta}$} & \text{$P$}  & \text{$\sigma$$\lambda$} & \text{BR} & \text{$\tau$} & \text{$B(M1)$} & \text{$B(E1)$} & \text{$B(E2)$} & \text{$B(E1)/B(E2)$}  \\

\text{} & (\text{keV}) &  &  &  & & &\text{(ps)} & $\mu_{N}^2$ & \text{($10^{-4}$ W.u.)} & (W.u.)  & (\text{$10^{-8}$$fm^{-2}$})\\

\hline

12$^{-}$  & 1487.3 &         & 0.71(14) & +0.98(62)&  $E1$	&   	 &  		        & 	 		        & 			 &		 	    &  \\

13$^{-}$  & 200.8  &         &  0.72(6) &          & $(M1/E2)$ &   	 &  		        & 	 		        & 			 &		 	    &  \\
	  & 1384.8 &         & 0.66(4)  & +0.47(21)& $E1$      &   	 &  	  		& 		 		&  		     	 &		 	    &  \\

14$^{-}$  & 126.7 & 0.88(13) & 1.23(14) &          & $M1/E2$   & 0.49(6) &  $<$ 4.2$^{b}$	& $>$ 1.52			& 			 &		 	    &  \\
	  & 941.6 &          & 0.51(8)  & +0.34(23)& $E1$      & 0.21(3) &  	  		& 		 		&  $>$ 0.21     	 &		 	    &  \\

15$^{-}$  & 140.4 & 0.83(25) & 1.08(6)  &          & $M1/E2$   & 0.89(8) &  $<$ 3.05$^{b}$	& $>$ 3.35			&  			 &		 	    &  \\
	  & 620.3 &          & 0.68(6)  & +0.22(15)& $E1$      & 0.11(2) &         		& 	 	 		& $>$ 0.54 		 &		 	    & 5.33 \\
	  & (266.2) & 	     & 	    	&	   & $(E2)$    & $<$ 0.01(1)&         		& 				&			 & $>$ 40.42	 	    &  \\

16$^{-}$  & 184.5 & 0.75(4)  & 0.84(10) & +0.01(15)& $M1/E2$   & 0.89(8) &  1.36$^{+0.22}_{-0.24}$ & 4.44$^{+1.14}_{-1.37}$ 	&  			 &		 	    &\\
	  & 646.6 & 0.87(9)  & 0.82(8)  & +0.47(32)& $E1$      & 0.10(2) &         		&  				& 0.98$^{+0.25}_{-0.26}$ &			    & 10.39$^{+5.21}_{-5.31}$ \\
	  & 325.0 & 	     & 	    	& 	   & $E2$      & 0.01(1)$^{c}$ &         		& 		 		&  			 & 37.62$^{+16.23}_{-16.45}$ &	\\

17$^{-}$  & 218.4 & 0.66(4)  & 0.81(9)  & +0.01(8) & $M1/E2$   & 0.91(9) &  1.18$^{+0.17}_{-0.18}$ & 	3.49$^{+1.85}_{-1.45}$	&  			 & 			    & \\
	  & 610.2 & 	     & 0.85(7)  & 	   & $E1$      & 0.07(2) &         		& 				& 0.94$^{+0.30}_{-0.30}$ &			    & 12.48$^{+7.62}_{-7.64}$\\
	  & 401.8 & 	     & 	    	& 	   & $E2$      & 0.02(1) &         		& 				&  			 & 30.03$^{+15.62}_{-15.70}$& \\

18$^{-}$  & 394.3 & 0.52(4)  & 0.68(7)  & +0.01(5) & $M1/E2$   & 0.90(9) &  0.89$^{+0.15}_{-0.13}$ & 	0.94$^{+0.18}_{-0.17}$	&  			 &			    &\\
	  & 491.0 & 	     & 0.68(12) & 	   & $E1$      & 0.05(1) &         		& 				& 1.70$^{+0.44}_{-0.42}$ &			    & 55.61$^{+20.47}_{-19.46}$\\
	  & 611.5 & 	     & 	    	& 	   & $E2$      & 0.05(1) &         		& 				&			 & 12.19$^{+3.19}_{-3.02}$  &  \\

19$^{-}$  & 425.4 & 0.53(6)  & 0.68(5)  & -0.12(8) & $M1/E2$   & 0.86(10)&  1.17$^{+0.50}_{-0.35}$ & 	0.54$^{+0.24}_{-0.17}$	&  			 &			    &\\
	  & 539.1 & 	     & 0.86(10) & 	   & $E1$      & 0.08(2) &         		& 				& 1.56$^{+0.77}_{-0.61}$ & 			    & 242.98$^{+163.45}_{-126.50}$\\
	  & 820.2 &          & 2.03(22) &          & $E2$      & 0.06(1) &         		& 				&			 & 2.56$^{+1.17}_{-0.88}$   &  \\

20$^{-}$  & 540.7 & 0.55(8)  & 0.97(10) & -0.01(4) & $M1/E2$   & 0.92(24)&  1.10$^{+0.17}_{-0.15}$ & 	0.30$^{+0.09}_{-0.09}$	&  			 &			    & \\
	  & 967.0 &          & 1.41(18) &          & $E2$      & 0.08(3) &         		& 				& 			 & 1.60$^{+0.65}_{-0.64}$   & \\

21$^{(-)}$& 562.0 &          & 1.08(19) &          & $(M1/E2)$ & 0.80(11)&  $<$ 0.46$^{b}$	& $>$ 0.56			& 			 &			    & \\
	  & 1103.0&          &          &          & $(E2)$    & 0.20(4) & 	       		& 				&			 & $>$ 4.94		    &  \\

\midrule
\midrule

\end{tabular}
\end{ruledtabular}
\footnotetext{$^{a}$uncertainty in $\gamma$ ray energy is $\pm$ (0.1-0.3) keV.}
\footnotetext{$^{b}$effective lifetime}
\footnotetext{$^{c}$actual error is $\sim$ 40\% which is rounded of to 100\% for two decimal place.}
\end{table*}

In the $^{142}$Eu nucleus, M. Piiparinen \textit{et al.} \cite{piipar} had identified two structures; one has positive parity whereas other one is of negative parity. The negative parity structure was found to populate to the several levels of the positive parity states via the $\gamma$ transitions among which most of them have $E1$ multipolarity. At lower spin positive parity states were found favourable whereas negative parity states were started to exhibit in the level structure at and above the $I^{\pi}$ = 12$^{-}$. Though, Piiparinen \textit{et al.} had explored the positive parity band (dipole) structure above 13$^{+}$ state the lower excited states were scared of such structural sequence. We have rearranged the $\gamma$ transitions de-exciting these states and a band-like structure, labelled as DB I, starting from 10$^{+}$ to 21$^{+}$ has been established as shown in Fig. \ref{levelsc}. Two newly observed $\gamma$-ray transitions of energy 620.3- and 413.0-keV have been placed confirmly from the present coincidence analysis in between the 15$^{+}$ $\rightarrow$ 13$^{+}$ and the 16$^{+}$ $\rightarrow$ 14$^{+}$ states, respectively, of DB I in $^{142}$Eu. Measured ADO ratio (R$_{\theta}$) for 413.0-keV transition was found to be 1.89(0.21) ($\sim$ 1.6 for pure quadrupole) which was in-agreement with the $E2$ character. For the weak nature of 620.3-keV results the tentative multipoarity assignment of the transition.

The previously observed dipole transitions of energy 126.7, 140.4, 184.5, 218.4, 394.3, 425.4, 540.7 and 562.0-keV in the negative parity sequence have been confirmed in the present study and therefore, the band-like structure has been assigned as DB II (Fig. \ref{levelsc}). The present coincidence measurements also confirm four new weak cross-over 266.2, 325.0, 401.8 and 611.5-keV $E2$ transitions between  15$^{-}$ $\rightarrow$ 13$^{-}$, 16$^{-}$ $\rightarrow$ 14$^{-}$, 17$^{-}$ $\rightarrow$ 15$^{-}$ and 18$^{-}$ $\rightarrow$ 16$^{-}$ states, respectively, in DB II. The $R_{\theta}$ values for 325.0- and 401.8-keV transitions were found to be of 1.30(21) and 1.43(22) respectively, which were well agreement with the $E2$ nature of these cross-over transitions. M. Piiparinen \textit{et al.} had observed five $E1$ 1487.2, 1384.8, 941.6, 620.3, 646.6-keV transitions through which the negative parity dipole band DB II decays to the positive parity band DB I \cite{piipar}. In the present work a survey of these transitions has been performed and subsequent confirmation has been made through the spectroscopic measurements. Moreover, four new connecting transitions of energy 610.2, 491.0, 539.1 and 295.6-keV have been observed for higher spin  states between the aforementioned bands from the coincidence analysis and placed them accordingly (Fig. \ref{levelsc}). The R$_{\theta}$ values for these transition are 0.73(5), 0.68(7), 0.59(6) and 0.52(10) respectively, confirmed their $E1$ character. Measured R$_{DCO}$ and $P$ values of these transitions also in support of their $E1$ nature.

\begin{figure}
	\centering
\setlength{\unitlength}{0.05\textwidth}
\begin{picture}(10,7)
\put(-0.4,-0.4){\includegraphics[width=0.54\textwidth]{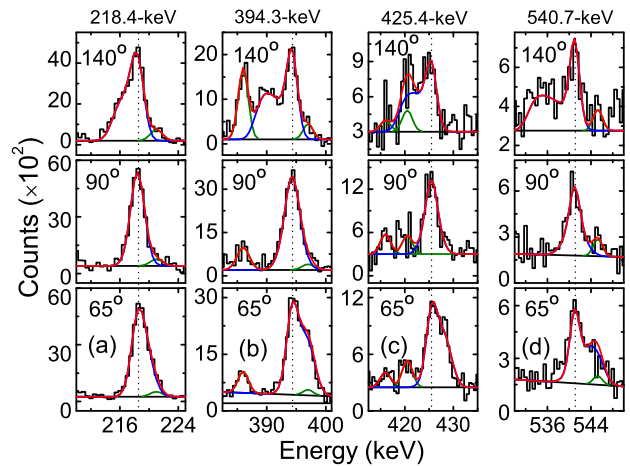}}

\end{picture}

\caption{\footnotesize (Color  online) The experimental line shapes fitted spectra for the (a) 218.4, (b) 394.3, (c) 425.4, and (d) 540.7-keV $\gamma$ transitions of the negative parity dipole band DB II in $^{142}$Eu. The first, second and last row corresponds to the spectra in the 140$^{\circ}$, 90$^{\circ}$ and 65$^{\circ}$, respectively. The blue, olive, and red curves represents the line shape of the $\gamma$ transition, contaminant peaks and total line shapes, respectively.}
	\label{shape}
\end{figure}

The level lifetime of the states in the bands DB I and DB II has been extracted by fitting the observed Doppler-broadened lineshapes of the de-exciting $\gamma$ transitions using the Doppler Shift Attenuation Method (DSAM). The LINESHAPE \cite{well-john, nrjonson2, lcnor} package has been used to simulate the trajectories of the residual nuclei ($^{142}$Eu) inside the target and backing medium. The simulated trajectories were then used to generate velocity profiles of the residual nuclei ($^{142}$Eu) for the clover detectors at different angles by assuming that the response of a composite clover detector was identical to a single crystal high-purity germanium (HPGe) detector with the dimension same as the dimension of the former placed at the same distance. The validity of this assumption was checked by analyzing the Doppler shapes obtained from a single crystal of a particular clover detector in the array \cite{rajban2} and comparing the lifetime results with that obtained from using the clover.

The level lifetimes have been obtained from the simultaneously fitting of experimental shape observed in the background-subtracted gated spectra of the angle dependent (65$^{\circ}$, 90$^{\circ}$ and 140$^{\circ}$) E$_\gamma$-E$_\gamma$ asymmetric matrices. The spectra at $90^{\circ}$ are very crucial for identifying the contaminant peak in the shape region. The gates have been set on the transitions below the band of interest which requires to consider the side-feeding contribution. The side feeding has been modeled with a cascade of five transitions having the same moment of inertia as that of the band under consideration \cite{nrjonson2}. Initially, starting from the topmost transition, the members of the band have been sequentially fitted. After having fitted all the transitions of the band, sequentially, a global least-squares minimization has been carried out for all the transitions of the cascade simultaneously, wherein only the transition quadrupole moments and the side-feeding quadrupole moments for each state have been kept as free parameters. To find out the effect of side-feeding on the evaluated lifetimes, we vary the side-feeding intensities between two extreme values (taking the corresponding uncertainties in intensities into account). The effect of variation in the side-feeding intensity resulted in a change in the level lifetime by less than 10\% \cite{rajban, sajad}. Finally, lifetimes were extracted by fitting the calculated line shape with the experimental spectra for minimum $\chi^2$ and  in the vicinity of this minimum value the uncertainties in the lifetimes are determined from the MINOS subroutine of the LINESHAPE program and it is not included systematic errors from the uncertainty in the stopping power, which can be as large as 15\%.

Doppler-broadened line shapes are observed for the transitions decaying from the states starting from the $I^\pi$ = 14$^-$ to the $I^\pi$ = 21$^-$ state of the negative parity dipole band in $^{142}$Eu. The 562.0-keV (21$^-$ $\rightarrow$ 20$^-$) transition is the highest transition of this dipole sequence for which a clear line-shape is observed in the experimental spectrum and an effective lifetime of 0.46 ps is extracted for the 21$^{-}$ state. This is then used as the input parameter for obtaining the level lifetimes of the lower lying states of the cascade. For the estimation of lifetime of all the state, the contribution of the observed side-feeding transitions are also taken into account. Typical fits to the observed Doppler shapes for the transitions of the negative parity dipole band in $^{142}$Eu are shown in Fig. \ref{shape}. The extracted level lifetimes and the corresponding reduced transition probabilities $B(M1)$, $B(E1)$, $B(E2)$ and $B(E1)/B(E2)$ values  have been depicted in Table \ref{table1}.

\begin{figure}
\centering
\setlength{\unitlength}{0.05\textwidth}
\begin{picture}(5,11)
\put(-1.8,0){\includegraphics[width=0.40\textwidth]{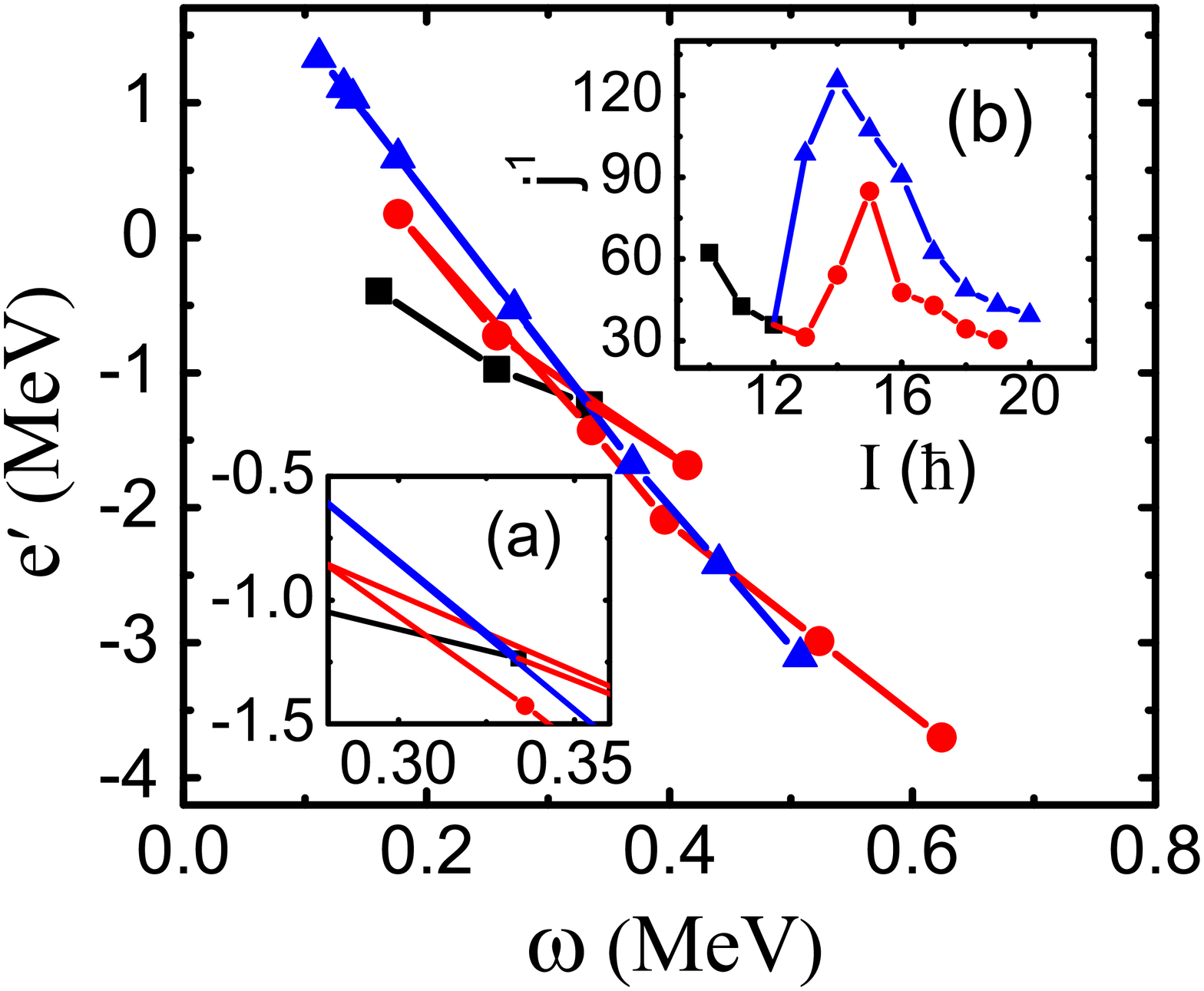}}
\put(-1.8,4.8){\includegraphics[width=0.40\textwidth]{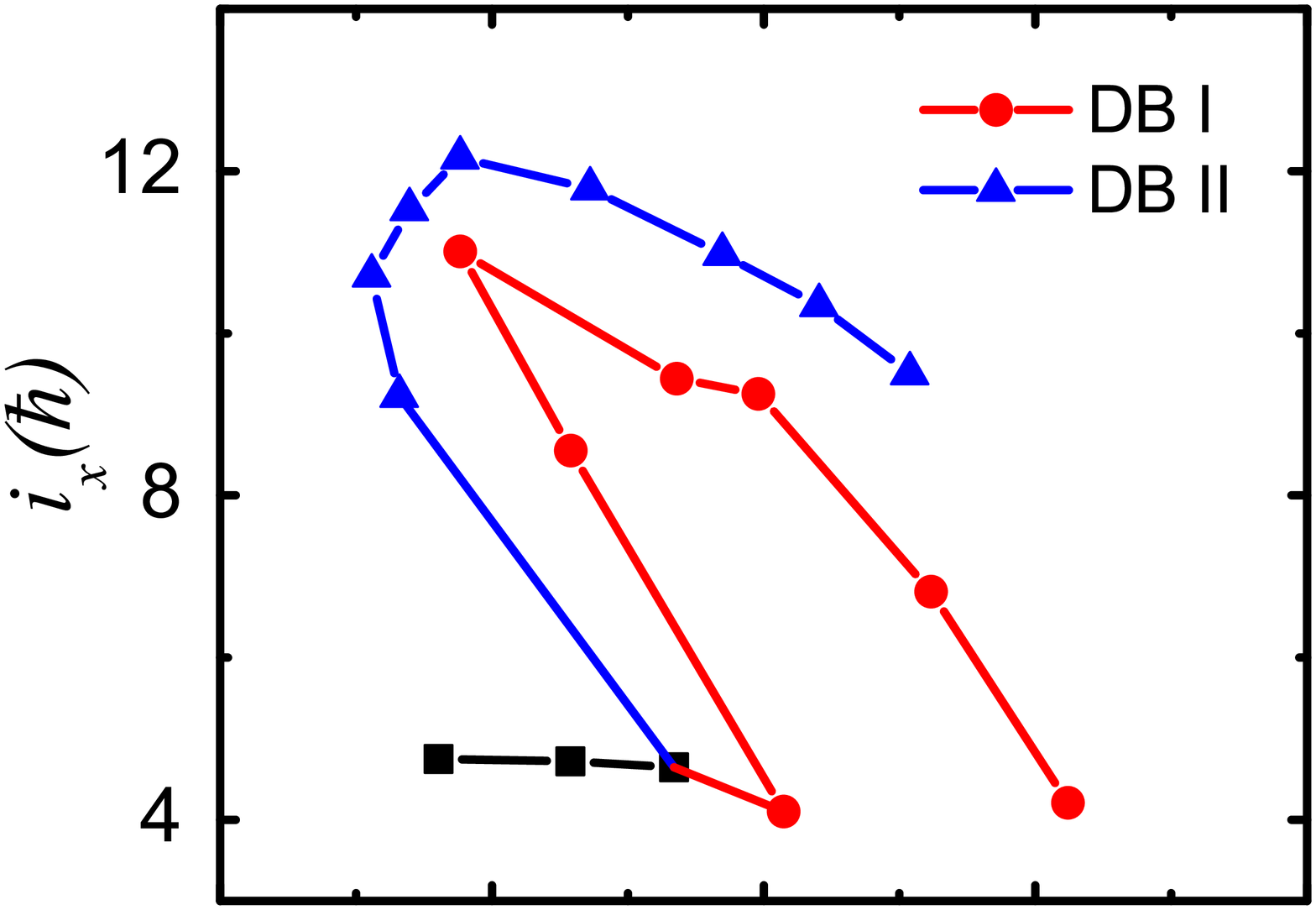}}

\end{picture}
\caption{\label{expt-char} The experimental characteristics extracted from the present investigation for the positive parity dipole band (DB I) and negative parity dipole band (DB II). Here, nature of the quasiparticle alignment ($i_{x}$) and the routhians (e$^{'}$) against the rotational frequency ($\omega$) of the states are shown in upper and lower panel respectively. The inset plot (a) inside lower panel shows zooming of the crossing region whereas inset plot (b) exhibit property of the kinetic moment of inerta ($j^{1}$) and with spin ($I$) of the states.}
\end{figure}

\begin{figure}
\centering
\setlength{\unitlength}{0.05\textwidth}
\begin{picture}(10,7.0)
\put(-0.1,-0.9){\includegraphics[width=0.55\textwidth, angle = 0]{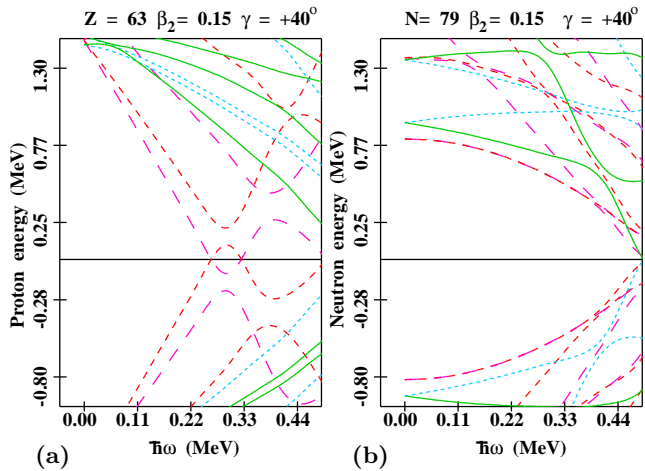}}
\put(0.5,0.0){\textbf{(a)}}
\put(5.2,0.0){\textbf{(b)}}
\end{picture}
\caption{\label{eng-orb} The energy behaviour of the orbitals against the rotational frequency ($\omega$) for (a) proton sector (b) neutron sector in $^{142}$Eu.}
\end{figure}

The excited states up to the 11$^{+}$ state in $^{142}$Eu was previously assigned as the member of the two quasiparticle configuration $\pi h_{11/2}^{1}$ $\otimes$ $\nu h_{11/2}^{-1}$. To assign the configurations of the observed bands DB I and DB II, we have calculated the kinetic moment of inertia ($j^{1}$), the quasiparticle alignment ($i_{x}$) and the experimental routhians (e$^{'}$) as shown in Figs. \ref{expt-char} (a) and (b) \cite{fraun}. It should be noted that both of the dipole bands DB I and DB II crosses at frequency $\sim$ 0.30 MeV and both of them have an alignment gain $\sim$ 8$\hbar$ with respect to the lower spin states, as shown in Figs. \ref{expt-char} (a) and (b). Such alignment may be due to the second crossing of $h_{11/2}$ orbital of the proton or the neutron as the first crossing is blocked in $^{142}$Eu. To understand the observed crossing we have calculated the neutron and the proton quasiparticle energies against the rotational frequency ($\omega$) at typical values of the deformation parameters $\beta_{2}$ = 0.15, $\beta_{4}$ = 0.02 and $\gamma$ = +40$^{\circ}$ as shown in Figs. \ref{eng-orb} (a) and (b). The calculations show that the neutron quasiparticle crossing for the $h_{11/2}$ orbital occurs at rotational energy ($\hbar\omega$) of $\sim$ 0.50 MeV while proton quasipaticle crosses at a rotational energy of $\sim$ 0.30 MeV. Therefore, observed experimental crossing at a rotational energy of $\sim$ 0.30 MeV may be due to the second crossing of the $h_{11/2}$ proton orbital for the bands. This conclusively establish the origin of the positive parity dipole band DB I as coupling of the two aligned $h_{11/2}$ proton quasiparticles with the two quasiparticles ($\pi h_{11/2}^{1}$ ${\otimes}$ $\nu{h}_{11/2}^{-1}$) configuration. This is consistent with the observed large alignment gain ($\sim 8\hbar$) at $\hbar$$\omega$ $\sim$ 0.30 MeV. Consequently, the configuration $\pi h_{11/2}^{3}$ ${\otimes}$ $\nu{h}_{11/2}^{-1}$ has been assigned to the dipole band DB I in $^{142}$Eu. The maximum spin that can be generated from this configuration is 19$\hbar$ which is in close agreement with the highest observed 21$^{+}$, 5533-keV state of this structure (DB I).

The observed negative parity states (DB II) above the 2283-keV 12$^{-}$ level ensure the coupling of odd number (at least one) of quasiparticles in the $h_{11/2}$ orbital with the $d_{5/2}$/$g_{7/2}$ orbital to generate the angular momentum of the states. It is evident from the calculated proton and neutron quasiparticles that the observed experimental alignment gain ($\sim$ 8$\hbar$) at the rotational frequency ($\hbar\omega$) $\sim$ 0.30 MeV (Fig. \ref{expt-char}) is due to the crossing of the protons $h_{11/2}$ orbital not for the neutrons in the $h_{11/2}$ orbital, as shown in the Figs. \ref{eng-orb} (a) and (b). For this reason, the $\pi h_{11/2}^{2} (g_{7/2}/d_{5/2})^{-1}$ ${\otimes}$ $\nu{h}_{11/2}^{-1}$ configuration, gives a maximum spin of 19$\hbar$, has been adopted for the negative parity band structure (DB II) in $^{142}$Eu.

\begin{figure}[t]
\centering
\setlength{\unitlength}{0.05\textwidth}
\begin{picture}(10,10.0)
\put(1.2,-0.0){\includegraphics[width=0.38\textwidth, angle = 0]{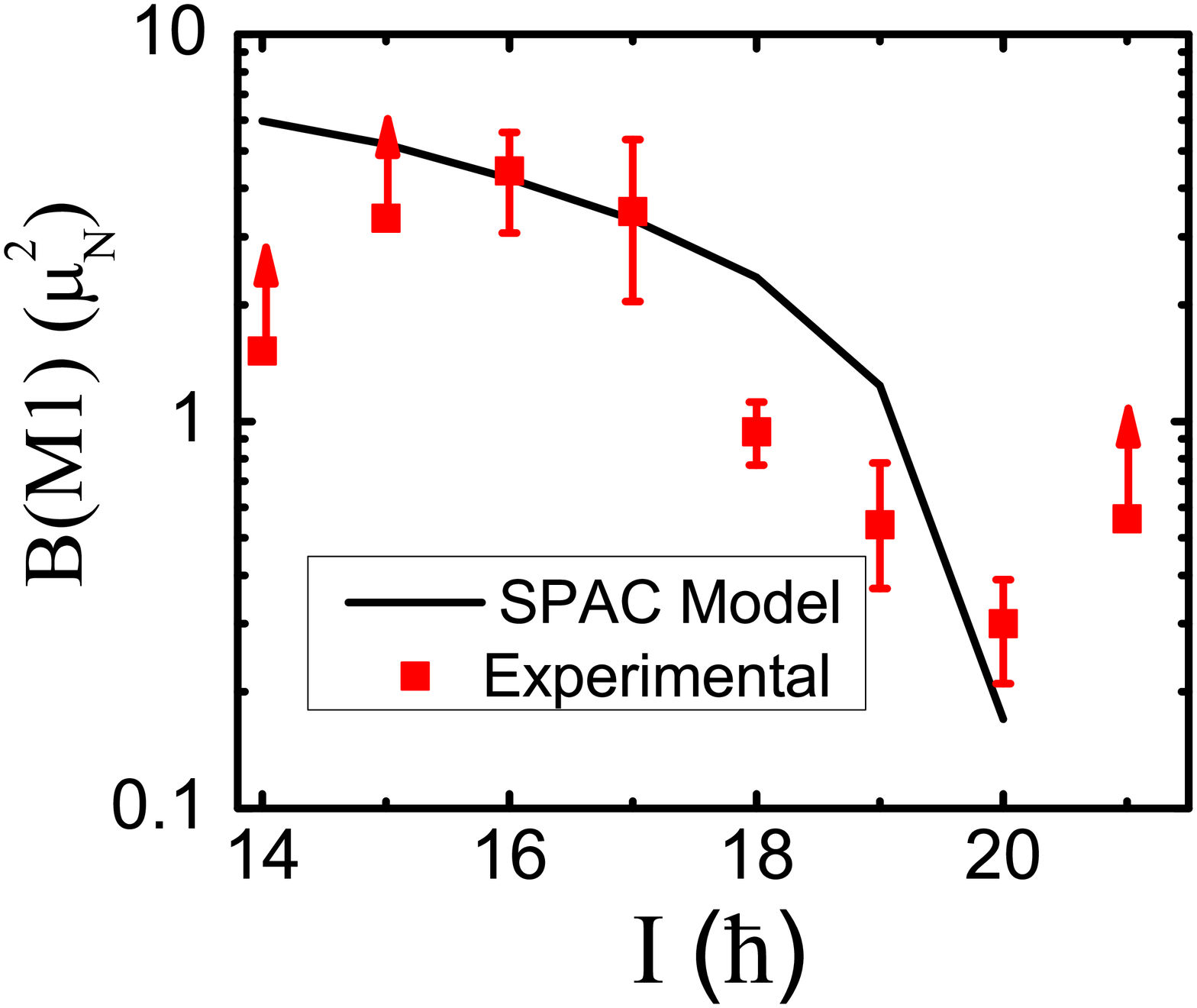}}
\put(1.2,+4.34){\includegraphics[width=0.38\textwidth, angle = 0]{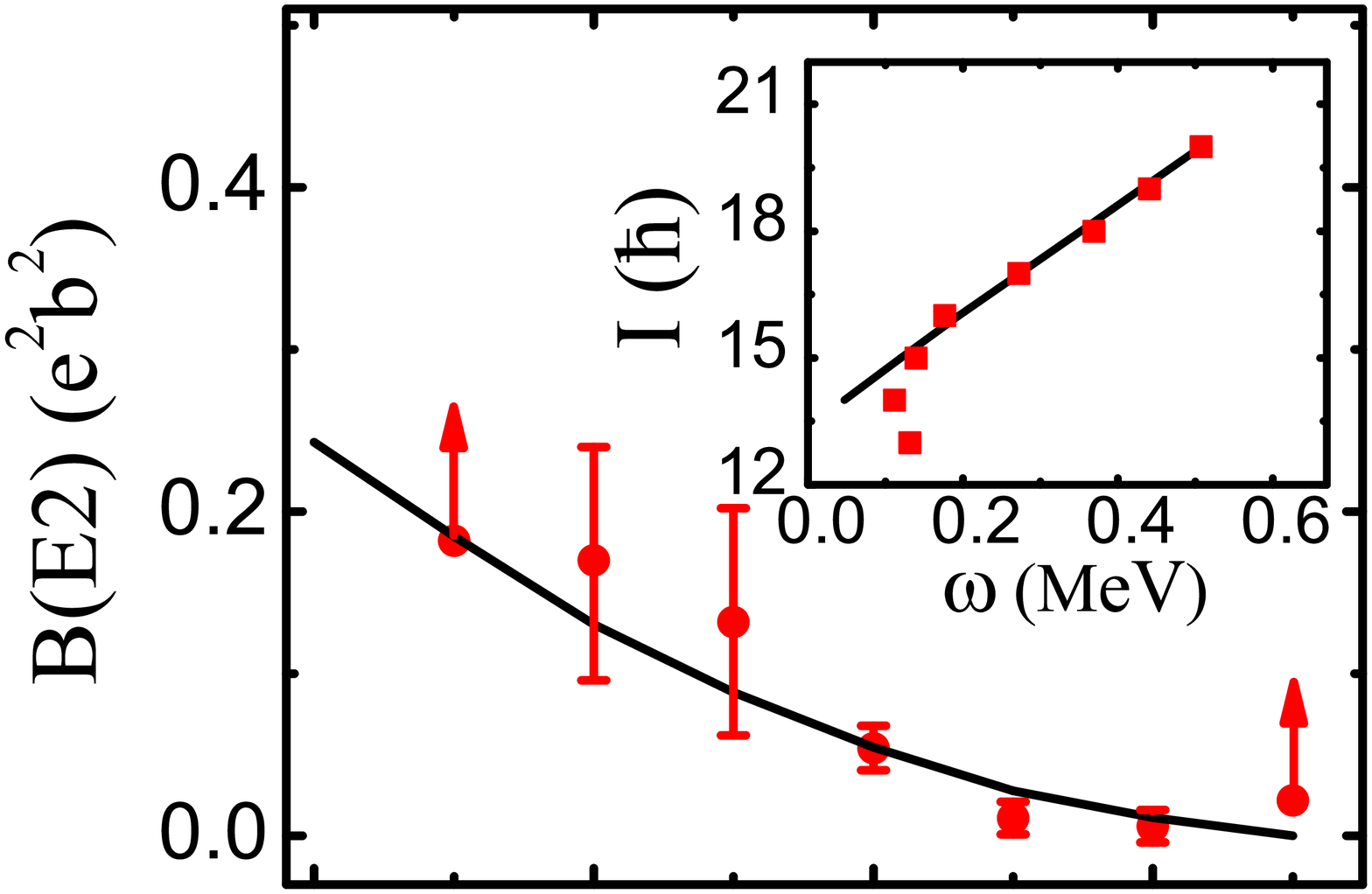}}
\put(3.2,6.1){\textbf{(b)}}
\put(3.2,1.4){\textbf{(a)}}
\end{picture}
\caption{\label{spac-model} Comparison of the experimental results (solid squares) of the DB III with the semiclassical SPAC model. The solid squares represents the experimental results for the DB IV. In panel (a) and (b) reduced transition strength $B(M1)$ and $B(E2)$ against spin ($I$) are plotted respectively. The inset plot in panel (b) shows variation of rotational frequency ($\hbar$) with angular momentum. The parameter used for the calculation are $j_1$ = 10$\hbar$, $j_{2}$ = 8$\hbar$, $g_1$ = -0.014, $g_2$ = 1.21, $J$ = 3.85$\hbar^2$/MeV and $v_2$ = 0.93 MeV, $Q_{eff}$ = 7.10, $Q_{col}$ = 0.50 \cite{rajban, sajad}.}
\end{figure}

The negative parity dipole band DB II consists with the dipole transitions of regular energy starting from spin $I^{\pi}$ = 13$^-$. Similar dipole bands in the neighboring $^{139, 141, 142}$Sm, $^{141, 143}$Eu, and $^{142}$Gd nuclei in this mass region ($A$ $\sim$ 140) have been labelled as MR bands form the decreasing nature of $B(M1)$ and $B(E2)$ values with spin \cite{139sm, sajad, rajban, pod, rajban2, past}. The evaluated $B(M1)$ and $B(E2)$ values for the DB II are also found to decrease with spin (Table \ref{table1} and Fig. \ref{spac-model}). This characteristic decrease of the transition rates (Fig. \ref{spac-model}) in DB II is hinted to the fact that the maximum contribution of the total angular momentum of the band DB II may be originated from the shears mechanism \cite{mac, clark}. To explore the intrinsic character of the band DB II the theoretical calculations within the framework of the shears mechanism with principle axis cranking model (SPAC) \cite{rajban, pod, past, sug} have been performed using the configuration $\pi{h_{11/2}^{2}} \otimes \pi{({{d_{5/2}/g_{7/2}})^{-1}}}\nu{h_{11/2}^{-1}}$. The SPAC model calculations have been elaborated in details in Refs. \cite{rajban,rajban2}. The total routhian surface calculations show that this configuration produces oblate deformation ($\beta_{2}$ $\sim$ -0.15) for the $^{142}$Eu nucleus. For oblate deformed shape and the normal initial alignment scheme of the SPAC model prefer the alignment of the holes and particles along the rotational axis and the symmetry (deformation) axis, respectively. Under these conditions, the energy of the states and the transition probabilities are calculated with particle and hole angular momenta, $j_{1}$ = 10$\hbar$ and $j_{2}$ = 8$\hbar$, respectively. Under these condition the experimental $B(M1)$ and $B(E2)$ values with spin and angular momentum with rotational frequency ($\omega$) are well reproduced within the SPAC model calculations as illustrated in Fig. \ref{spac-model}. The successful representation of the decreasing trend of experimental $B(M1)$ and $B(E2)$ values with spin as well as the behaviour of $\omega$ therein is strongly indicative of the fact that the maximum contribution to the angular momentum of the states constituting DB I is originated from the shears mechanism.

The negative parity structure DB II has been depopulated to the positive parity structure (DB I) through several strong electric dipole ($E1$) transitions having large $B(E1)$ rates as depicted in Table \ref{table1}. The estimated $B(E1)$ values are of the order of 10$^{-4}$ W.u. and $B(E1)/B(E2)$ values of the order of 10$^{-8}$ $fm^{-2}$, which is comparable to the values obtained in $^{124, 125}$Ba \cite{mason}, $^{117, 121}$Xe \cite{liu} and $^{124}$Cs \cite{selva}, where octupole correlation has been observed. Such a proposition would exist for the dipole band DB II which has been generated due to the promotion of the proton ($d_{5/2}$/$g_{7/2}$) particle to the $h_{11/2}$ orbital. This indicates that a dynamical octupole correlation between the $d_{5/2}$ and $h_{11/2}$ orbitals of protons near the Fermi surface in $^{142}$Eu would exists as evident from the observed strong $E1$ connecting transitions between the DB I and DB II. This is the first experimental observation of octupole correlation in weakly deformed nuclei throughout the nuclear chart. In the present case $E1$ transitions become weaker with increasing spin as evident from the Table \ref{table1} which imply the reduction of strength of the octupole correlation along the band DB II. This observation also leads to the unique observation of coupling between two extreme mode of generation of angular momentum i.e. shears mechanism and octupole correlation in a single nuclei.

In summary, the intrinsic structure of two opposite parity dipole bands DB I and DB II in $^{142}$Eu has been investigated using the fusion evaporation reaction $^{116}$Cd($^{31}$P,5n). The positive parity structure has been originated due to the alignment of the an extra pair of protons in $h_{11/2}$ with respect to the low spin structure (at and below the 11$^{+}$ state) as supported by the TRS calculations. This calculations give strong support of the promotion of an odd proton to the $h_{11/2}$ orbital from the ($g_{7/2}$/$d_{7/2}$) core for generation of the negative parity band DB II.  The dipole bands DB I and DB II have been assigned as the $\pi{h_{11/2}^{2}} \otimes \nu{h_{11/2}^{-1}}$ and $\pi{h_{11/2}^{2}} \otimes \pi{({{d_{5/2}/g_{7/2}})^{-1}}}\nu{h_{11/2}^{-1}}$ configurations, respectively. Theoretical calculations within the SPAC model framework are in well agreement with the experimental $\omega$, $B(M1)$ and $B(E2)$ transition rates reveal the MR phenomenon of the dipole band DB II. The shell model calculation exhibits the significant contribution from the proton $d_{5/2}$ orbital for the negative parity levels of DB II in $^{142}$Eu. The $E1$ transitions from the $\pi{h_{11/2}^{2}} \otimes \pi{({{d_{5/2}/g_{7/2}})^{-1}}}\nu{h_{11/2}^{-1}}$ band (DB II) to the $\pi{h_{11/2}^{3}} \otimes \nu{h_{11/2}^{-1}}$ band (DB I) have been observed in this nucleus, indicating that the corresponding orbitals $\pi{h_{11/2}}$ and $\pi{d_{5/2}}$ form an octupole driving pair. The deduced $B(E1)$ values indicate enhancement of octupole correlation in this nucleus similar to that reported for some of the Xe, Cs, Ba and Br isotopes \cite{liu, selva, mason, cliu}.

The authors gratefully acknowledge financial support from the Department of Science $\&$ Technology (DST) for the INGA Project (Project No. IR/S2/PF-03/2003-II). We would like to acknowledge help from all INGA collaborators. We thank uninterrupted $^{31}$P beam. G.G. acknowledges support provided by the University Grants Commission, Departmental Research Support (UGC-DRS) Program [Program No. F.530/16/DRS-II/2015(SAP-I)]. S. R. would like to acknowledge the financial assistance from the University Grants Commission - Minor Research Project (No. PSW-249/15-16 (ERO)).


\begin{thebibliography}{9}

\bibitem{fraun2} S. Frauendorf, Rev. Mod. Phys. {\bf 73}, 463 (2001).

\bibitem{mac} A.O. Macchiavelli, \textit{et al.}, Phys. Rev. C {\bf 57}, R1073  (1998).

\bibitem{clark} R.M. Clark, \textit{et al.}, Phys. Rev. Lett. {\bf 82}, 3220  (1999).

\bibitem{rev} P.A. Butler, W. Nazarewicz, Rev. Mod. Phys. {\bf 68}, 349 (1996).

\bibitem{139sm} A. A. Pasternak, \textit{et al.}, Eur. Phys. J. A {\bf 37}, 279–286 (2008).

\bibitem{sajad} S. Rajbanshi \textit{et al.}, Phys. Rev. C {\bf 94}, 044318 (2016).

\bibitem{rajban} S. Rajbanshi \textit{et al.}, Phys. Rev. C {\bf 89}, 014315 (2014).

\bibitem{pod} E. O. Podsvirova, \textit{et al.}, Eur. Phys. J. A {\bf 21} 1–6 (2004).

\bibitem{rajban2} S. Rajbanshi \textit{et al.}, Phys. Rev. C {\bf 90}, 024318 (2014).

\bibitem{past} A. A. Pasternak, \textit{et al.}, Eur. Phys. J. A {\bf 23} 191–196 (2005).

\bibitem{palit} R. Palit \textit{et al.}, Nucl. Instrum. Methods A {\bf 680}, 90 (2012).

\bibitem{htan} H. Tan \textit{et al.}, \textit{in Nuclear Science Symposium Conference Record 2008} (IEEE, Washington, DC, 2008), p. {\bf 3196}.

\bibitem{radford1} D. C. Radford, Nucl. Instrum. Methods, Phys. Res.,
 Sect {\bf A 361}, 297 (1995).

\bibitem{radford2} D. C. Radford, Nucl. Instrum. Methods, Phys. Res., Sect {\bf A 361}, 306 (1995). 

\bibitem{ingasort} R. K. Bhowmik, INGASORT manual (private communication).


\bibitem{piipar} M. Piiparinen \textit{et al.}, Nucl. Phys. A {\bf 605}, 191 - 268 (1996).

\bibitem{well-john} J. C. Wells and N. R. Johnson, LINESHAPE: A Computer Program
for Doppler Broadened Lineshape Analysis, Report No. ORNL-
6689, 44, 1991.

\bibitem{nrjonson2} N. R. Johnson, \textit{et al.}, Phys. Rev. C {\bf 55}, 652 (1997).

\bibitem{lcnor} L. C. Northcliffe and R. F. Schilling, Nucl. Data Tables A {\bf 7}, 233 (1970).

\bibitem{fraun} R. Bengtsson and S. Frauendorf, Nuclear Physics A {\bf 327},139-171 (1979).

\bibitem{sug} M. Sugawara et al., \textit{et al.}, Phys. Rev. C {\bf 79}, 064321 (2009).

\bibitem{mason} P. Mason \textit{et al.}, Phys. Rev. C {\bf 72}, 064315 (2005).

\bibitem{liu} Z. Liu \textit{et al.} Eur. Phy. J. A {\bf 1}, 125 (1998).

\bibitem{selva} K. Selvakumar \textit{et al.}, Phys. Rev. C {\bf 92}, 064307 (2015)

\bibitem{cliu} C. Liu \textit{et al.}, Phys. Rev. Lett. {\bf 116}, 112501 (2016)



\end{thebibliography}
\end{document}